%% file: main.tex
\documentclass[%
 reprint,
superscriptaddress,
 amsmath,amssymb,
 aps,
 prl,
]{revtex4-1}
\usepackage{graphicx}
\usepackage{siunitx}
\usepackage{xcolor}
\usepackage[utf8x]{inputenc}
\usepackage{ulem}
\usepackage{supertabular}
\usepackage{slashed}
\usepackage{hyperref}

\definecolor{gesfblack}{rgb}{0,0,0}

\definecolor{gesfblue}{rgb}{0.08,0.42,0.76}

\definecolor{gesfgreen}{rgb}{0,1,0}

\definecolor{gesfgrey}{rgb}{0.5,0.5,0.5}

\definecolor{gesflanse}{rgb}{0.00,0.50,0.50}

\definecolor{gesfpurple}{rgb}{0.47,0.19,0.42}

\definecolor{gesfred}{rgb}{1,0,0}

\definecolor{gesfwhite}{rgb}{1,1,1}

\definecolor{gesfyellow}{rgb}{0.7,0.4,0.3}

\begin{document}
\title{Search for Cosmic-ray Boosted Sub-MeV Dark-Matter-Electron Scattering in PandaX-4T} 

\input{authorlist}

\date{\today}

\begin{abstract}

We report the first search for the elastic scatterings between cosmic-ray boosted sub-MeV dark matter and electrons in the PandaX-4T liquid xenon experiment. Sub-MeV dark matter particles can be accelerated by scattering with electrons in the cosmic rays and produce detectable electron recoil signals in the detector. Using the commissioning data from PandaX-4T of 0.63~tonne$\cdot$year exposure, we set new constraints on DM-electron scattering cross sections for DM masses ranging from 10~eV/$c^2$ to 3~keV/$c^2$. 
\end{abstract}

\maketitle

The nature of dark matter (DM) is still a major mystery in modern physics~\cite{evidence2005,evidence2018}. In particular, DM mass remains unknown and its possible values span tens of orders in magnitude. Direct detection experiments usually focus on DM mass above GeV for the traditional weakly-interacting massive particle with nuclear recoils (NRs)~\cite{PandaX-4T:2021bab,XENON:2023cxc,LZ:2022lsv}, and can go lower to MeV range with electron recoils (ERs)~\cite{PandaX:2022xqx,DAMIC-M:2023gxo,SENSEI_2023,DarkSide:2022knj}. However, a few very well-motivated DM models, such as the freeze-in~\cite{Bernal:2017kxu,Hall:2009bx} or the asymmetric DM~\cite{Kaplan:2009ag,Zurek:2013wia,Petraki:2013wwa}, with a characteristic DM mass in the sub-MeV region, remain much less explored due to the detection threshold in the direct detection experiments. 
In recent years, direct detection experiments have also begun to apply new mechanisms, such as various acceleration processes~\cite{Bringmann2018,CReDM_2018,Ding_CR,Shujing_CReDM,LiJinmian,2024_Atanu,Xia:2024ryt,JinWei_blazer_e,boosted_Sun_An,CDEXsolar} and absorption~\cite{Yu_absorption,Ge:2022ius,PandaX:2022ood}, in order to surpass the detector threshold to probe sub-MeV DM.

Previously, experiments searching for the scatterings between nucleons and sub-GeV DM accelerated by cosmic rays (CRs) have been carried out by PROSPECT~\cite{2021PROSPECT}, PandaX~\cite{PandaX-II:2021kai}, CDEX~\cite{2022CDEX} and SuperK\cite{2023SuperK}.
Similarly, if DM particles can scatter with electrons in the detector, they will inevitably collide with electrons in the cosmic rays. This collision can boost their kinetic energy. Consequently, in the detector, the electron recoil energy obtained from the scattering by the DM can exceed the detection threshold even for sub-MeV DMs. In this letter, we report a search for the ER signals from the scatterings of cosmic ray electron boosted DMs (CReDM) in the PandaX-4T liquid xenon detector~\cite{PandaX-4T:2021bab}, located in the China Jinping Underground Laboratory (CJPL)~\cite{CJPL1,CJPL2}. 

The prediction of CReDM signals involves the acceleration of DM by CRs,  the attenuation of DM flux in the Earth, and the DM scattering with electrons in the detector.
For the acceleration process, we employ the approach in Ref.~\cite{Xia:2022tid}, which simulates the Galactic CR electron distribution using the \texttt{GALPROP} code with the best-fit parameters from the \texttt{GALPROP-HelMod} analysis~\cite{Boschini:2018zdv}, and assumes a Navarro-Frenk-White halo profile~\cite{Navarro:1996gj} for the Galactic DM distribution. The local density of the DM is set to be the conventional value of $\rho_{\chi} = 0.3$~GeV/cm$^{3}$~\cite{PandaX-4T:2021bab}.
We consider a typical scenario where the fermionic DM particle $\chi$ and the electron $e$ interact in the form $\mathcal{L}_{\rm int} = G \bar{\chi} \gamma^\mu \chi \bar{e} \gamma_\mu e$, where $G$ is the effective vector coupling constant. The vector coupling is realized in popular models such as dark photon~\cite{Fabbrichesi:2020wbt} and $B - L$ gauge boson models~\cite{BL_model}.
The differential scattering cross section for the relativistic CR electron scattering off the halo DM, which is approximately at rest, can be calculated as
\begin{equation}
\begin{aligned}
  \frac{d\sigma_{\chi e}}{dT_\chi}
= 
  & \frac{\bar{\sigma}_{\chi e}}{4\mu_{\chi e}^2 T_e^{\rm CR} (T_e^{\rm CR} + 2m_e)}  
  \Big[
  2m_\chi(m_e + T_e^{\rm CR})^2
  \\
  & - T_\chi \left(2m_\chi T_e^{\rm CR} + (m_\chi + m_e)^2\right)
  +
  m_\chi T_\chi^2
  \Big],
\end{aligned}
\label{eq:diff_flux}
\end{equation}
where $m_\chi$ ($m_e$) and $T_\chi$ ($T_e^{\rm CR}$) are the mass and kinetic energy of the DM particle (CR electron), respectively, $\mu_{\chi e} \equiv m_\chi m_e / (m_\chi + m_e)$ is the DM-electron reduced mass, and the effective cross section $\bar{\sigma}_{\chi e}$ is defined as
$\bar{\sigma}_{\chi e} \equiv G^2 \mu_{\chi e}^2/\pi$. 

CJPL has a rock overburden of about 2.4~km.
To calculate the attenuation of the CReDM, we employ the Monte Carlo (MC) simulation method developed in our previous work~\cite{PandaX-II:2021kai}, with nucleon targets replaced by electrons. Although electrons within the Earth are usually bound within atoms, the mean energy transfer associated with the attenuation process for a typical MeV-energy CReDM particle is around 0.3 MeV, which is significantly larger than the binding energy of atoms which is $\mathcal{O}(10)$ keV at most. Therefore, we employ free electron scattering in our simulation for simplicity. The resultant differential scattering cross section has the same formula as Eq.~\ref{eq:diff_flux} with  $\chi$ and $e$ exchanged. Since the scattering cross section for free electrons is typically larger than that for bound electrons~\cite{Ge:2021snv}, this approximation results in stronger attenuation and, consequently, more conservative DM signal yield in the detector.

\begin{figure}[htbp]
  \includegraphics[width=0.48\textwidth]{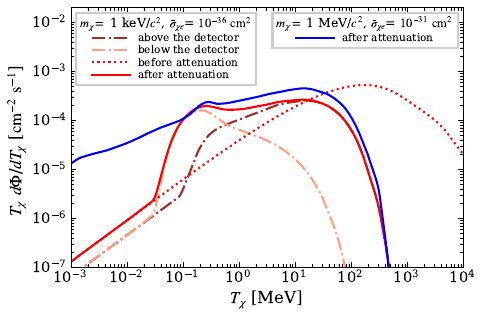}
  \includegraphics[width=0.48\textwidth]{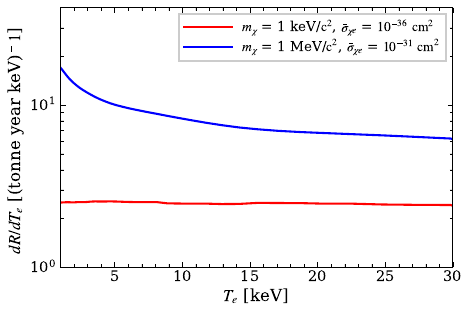}
  \caption{The upper panel shows the flux of CReDM for $m_\chi = 1~\text{keV}$/$c^2$ and  $\bar{\sigma}_{\chi e} = 10^{-36}~\text{cm}^2$ on the Earth's surface and that reaching the PandaX-4T detector after attenuation, including particles that
  come from above and below the detector's horizontal plane. For comparison, the CReDM flux after attenuation with $m_\chi = 1~\text{MeV}$/$c^2$ and  $\bar{\sigma}_{\chi e} = 10^{-31}~\text{cm}^2$ is also shown in the blue line. The lower panel shows the differential recoil events rate in the xenon detector for $m_\chi$ = 1~keV/$c^2$ and $m_\chi$ = 1~MeV/$c^2$.}
\label{fig:flux_CReDM}
\end{figure}

The upper panel of Fig.~\ref{fig:flux_CReDM} shows the differential CReDM flux reaching the detector ${d\Phi}/{dT_\chi}$ for a set of benchmark parameters $m_\chi = 1~\text{keV}$/$c^2$ and $\bar{\sigma}_{\chi e} = 10^{-36}~\text{cm}^2$, compared with the surface flux without the Earth attenuation. As we can see, most DM particles after acceleration become relativistic. The flux is distorted and shifted from high energy to low energy after attenuation. The two bumps are contributed by DM particles 
coming from above and below the detector's horizontal plane.
At DM energy below $\sim 3\times 10^{-2}~\text{MeV}$, the underground flux is nearly identical to that at the surface.
This is because the energy dependent cross section $\sigma_{\chi e}$ of DM scattering with electrons in the Earth decreases as $T_\chi$ decreases. At $T_\chi = 3\times 10^{-2}~\text{MeV}$, the corresponding mean free path is more than twice of the diameter of the Earth. 

For the scattering of DM with electrons in the liquid xenon detector, we take into account the effect of bound electrons, since the detection threshold ($\sim$ keV) is comparable to the binding energy of 
the inner electrons in xenon atoms. The differential cross section for DM particle scattering off bound electrons in the shell with the principal (angular momentum) quantum number $n(l)$, can be calculated as 
\begin{equation}
\begin{aligned}
    \frac{d\sigma^{nl}_{\chi e}}{dT_e}
=
    \frac{\bar{\sigma}_{e\chi}}{8\mu_{e\chi}^2 v_\chi^2 T_e}
    \int_{|\mathbf{q}|_{\min}}^{|\mathbf{q}|_{\max}} d |\mathbf{q}|\,
    |\mathbf{q}|\, |f^{nl}_{\rm ion}(T_e, |\mathbf{q}|)|^2
    \\
    \times
    \left(1 - \frac{\Delta E_{nl}}{E_\chi} - \frac{\mathbf{q}^2 - \Delta E_{nl}^2}{4E_\chi^2}\right),
\end{aligned}
\label{eq:xsec_bound}
\end{equation}
where $T_e$ is the recoil energy of the electron, $v_\chi = |\mathbf{p_\chi}| / E_\chi$ is the speed of the incident DM particle, the energy transfer $\Delta E_{nl}$ is the sum of $T_e$ and the binding energy of the $(n,l)$ shell, $\mathbf{q}$ is the three-momentum transfer, the integration limits  $|\mathbf{q}|_{\min / \max}(T_\chi, T_e)$ are given by $|\mathbf{p_\chi}| \mp \sqrt{(E_\chi - \Delta E_{nl})^2 - m_\chi^2}$,  and $|f^{nl}_{\rm ion}(T_e, |\mathbf{q}|)|^2$ is the ionization factor calculated by the \texttt{DarkART} code~\cite{Catena:2019gfa,DarkART_zenodo} which treats the electron as non-relativistic. In the derivation of Eq.~\ref{eq:xsec_bound}, the electron is considered as non-relativistic to be consistent with the ionization factor, while no approximation is made for the DM (see Supplemental Material for details). It is worthy to note that, for the typical energy $E_\chi \sim 1$~MeV of CReDM, the factor in the second line of Eq.~\ref{eq:xsec_bound} is close to 1 in the dominant region of $|\mathbf{q}|$, and the differential cross section becomes independent of $T_e$
because the integral turns out to be approximately proportional to $T_e$, effectively canceling the $T_e$ in the prefactor~\cite{Ge:2021snv}. 

Then the differential electron recoil events rate can be calculated as
\begin{equation}
    \frac{dR}{dT_e}
=
    N_T
    \sum_{nl}
    \int_{T_\chi^{\min}(T_e)}^{\infty} dT_\chi
    \frac{d\sigma^{nl}_{\chi e}}{dT_e}
    \frac{d\Phi}{dT_\chi}
    \;,
\label{eq:dRdTe}
\end{equation}
where $N_T\approx 4.6\times10^{24}/{\rm kg}$ is the number of xenon atoms per unit target mass, $T_\chi^{\min}$ = $\Delta E_{nl}$ is the minimal kinetic energy of DM particle to produce an electron recoil energy of $T_e$,  
and the sum is over electron shells from $1s$ to $5p$.
The lower panel of Fig.~\ref{fig:flux_CReDM} shows the differential event rate for two typical DM masses, 1 keV/$c^2$ and 1 MeV/$c^2$.
For the 1 keV/$c^2$ mass case (much less than typical $T_\chi$ of $\sim$MeV), the recoil event rate exhibits a flat spectrum in the energy region of interest for this analysis, for two reasons.
First, the differential cross section  $d\sigma^{nl}_{\chi e}/{dT_e}$ in Eq.\ref{eq:dRdTe} is approximately independent of $T_e$, as discussed below Eq.\eqref{eq:xsec_bound}.
Second, since the $v_\chi^{-2}$ factor in $d\sigma^{nl}_{\chi e}/{dT_e}$ is $\sim 1$ for keV/$c^2$ mass DM, then the $T_\chi$ integration is mainly contributed by the CReDM flux $d\Phi/dT_\chi$ at higher energies around MeV (as shown in the upper panel of Fig.~\ref{fig:flux_CReDM}). Thus the $T_e$ dependence of the lower limit $T_\chi^{\min}\sim T_e$ at keV scale does not significantly impact the integration result, leading to an almost constant $dR/dT_e$.
For the 1~MeV/$c^2$ mass case, the DM particle is non-relativistic near the threshold energy, thus the $v_\chi^{-2}$ factor is proportional to $T_\chi^{-1}$ that enhances the contribution from the low-energy part of the CReDM flux $d\Phi/dT_\chi$, resulting in an increase in the recoil event rate toward the low energy.

PandaX-4T utilizes a dual-phase xenon time projection chamber (TPC) which contains 3.7 tonne of liquid xenon in the sensitive volume. When a particle scatters with xenon nucleus or electron in the liquid xenon, prompt scintillation photons ($S1$) and delayed proportional electroluminescence photons ($S2$) are generated. Both signals are collected by top and bottom arrays of 368 Hamamatsu R11410-23 3-inch photomultiplier tubes (PMTs).  
The $S1$ and $S2$ signal distributions from NR and ER in the PandaX-4T detector are simulated using the standard NEST v2.2.1 framework~\cite{NESTv2.2.1,Szydagis:2021hfh} with parameters fitted to low energy calibration data. More detailed descriptions of the PandaX-4T experiment can be found in Ref.~\cite{PandaX-4T:2021bab}.

This analysis uses the data from PandaX-4T commissioning run, corresponding to an exposure of 86.0 live-days. The data quality selection criteria are the same as in  Ref.~\cite{PandaX-4T:2021bab}. The electron-equivalent energy of a given event is reconstructed as $E_{\rm{ee}}= 13.7~{\rm eV}\times(S1/{\rm{PDE}} + S2_b/{\rm{EEE}}/{\rm{SEG_b}})$, where $S2_b$ is the bottom-only $S2$ to avoid PMT saturations. PDE, EEE, and SEG$_b$ represent the photon detection efficiency, electron extraction efficiency, and single electron gain using $S2_b$, respectively. Their values are taken from Ref.~\cite{PandaX-4T:2021bab}. The lower energy threshold of the DM candidates is defined by requiring a $S1$ above 2 photoelectrons (PEs) and a $S2>$80 PEs. To cover more CReDM signals, we extend the energy range up to 30~keV while avoiding the backgrounds from $^{124}$Xe, $^{125}$I and $^{127}$Xe in the range of 30 to 40 keV. In total, 1116 events are selected in the data. 

In this analysis, relevant backgrounds, including tritium, material radioactivity, ${}^{8}$B neutrinos, neutrons, ${}^{127}$Xe, surface events and accidental background are estimated according to the method in the previous analysis~\cite{PandaX-4T:2021bab} and our extended energy region. Other backgrounds including $^{222}$Rn, $^{220}$Rn, ${}^{85}$Kr, have been better understood and re-estimated in Ref.~\cite{Xe134_yxy}. Their nominal contributions and uncertainties in our energy window are summarized in Table~\ref{tab:1}. The assumed spectra of $^{222}$Rn, $^{220}$Rn, ${}^{85}$Kr are the same as in our previous analysis in Ref.~\cite{solar_PP}.  

Solar neutrino background mainly includes solar $\mathit{pp}$ neutrino from proton-proton fusion chain and solar ${}^{7}$Be neutrino from electron capture of ${}^{7}$Be according to the Standard Solar Model~\cite{SSM}. Their contributions are estimated using the energy spectrum in Ref.~\cite{Chen:2016eab}, with xenon atomic effects taken into account, resulting in $43.3\pm 8.9$ events. 

The background from ${}^{136}$Xe two-neutrino double-$\beta$-decay is estimated to be $34.0 \pm 1.9$ events, using the half-life measured in the PandaX-4T experiment~\cite{Xe136_silin}. The process of two-neutrino double-electron-capture of ${}^{124}$Xe, which has been observed by XENON1T experiment~\cite{XENON_xe124_nature}, deposits an energy around 10~keV (LL shells). Its contribution is estimated to be $2.2 \pm 0.5$ events according to the measured half-life~\cite{XENON_xe124}.

\begin{table}[htbp]
    \centering
    \setlength{\tabcolsep}{3.0mm}{
    \begin{tabular}{c c c}
    \toprule
        component & Expected counts & Best-fit counts  \\
        \hline
        Tritium &$525 \pm 31$ & $525 \pm 31$  \\
        $^{222}$Rn & $ 323\pm 16$  & $321 \pm 13$   \\
        $^{220}$Rn & $ 58 \pm 15$ &  $57\pm 13$  \\
        $^{85}$Kr & $ 94\pm 47$  &  $81 \pm 24$  \\
        Material & $ 41.9\pm 8.6$  & $41.4 \pm 8.0$ \\
        Solar neutrino & $43.3\pm 8.9$  & $42.7 \pm 8.3$ \\
        $^{136}$Xe & $34.0 \pm 1.9$  & $33.8 \pm 1.9$ \\
        $^{127}$Xe & $8.2 \pm 2.1$  & $8.4 \pm 2.0$  \\
        $^{124}$Xe & $2.2 \pm 0.5$  & $2.2 \pm 0.5$ \\
        $^{8}$B & $0.8 \pm 0.3$  & $0.8 \pm 0.2$ \\
        Neutron & $1.1 \pm 0.6$  & $1.1 \pm 0.5$ \\
        Accidental & $2.4 \pm 0.5$  & $2.4 \pm 0.6$ \\
        Surface & $0.5 \pm 0.1$  & $0.5 \pm 0.1$ \\
        \hline
        Sum & $1134 \pm 75$  & $ 1117\pm 50$  \\  
        \hline
        Data & 1116  \\
    \botrule
    \end{tabular}}
    \caption{Expected background contribution of each kind in selected data, as well as the background-only best-fit values. The tritium values are obtained from unconstrained fit.}
    \label{tab:1}
\end{table}

\begin{figure}[htbp]
  \includegraphics[width=0.48\textwidth]{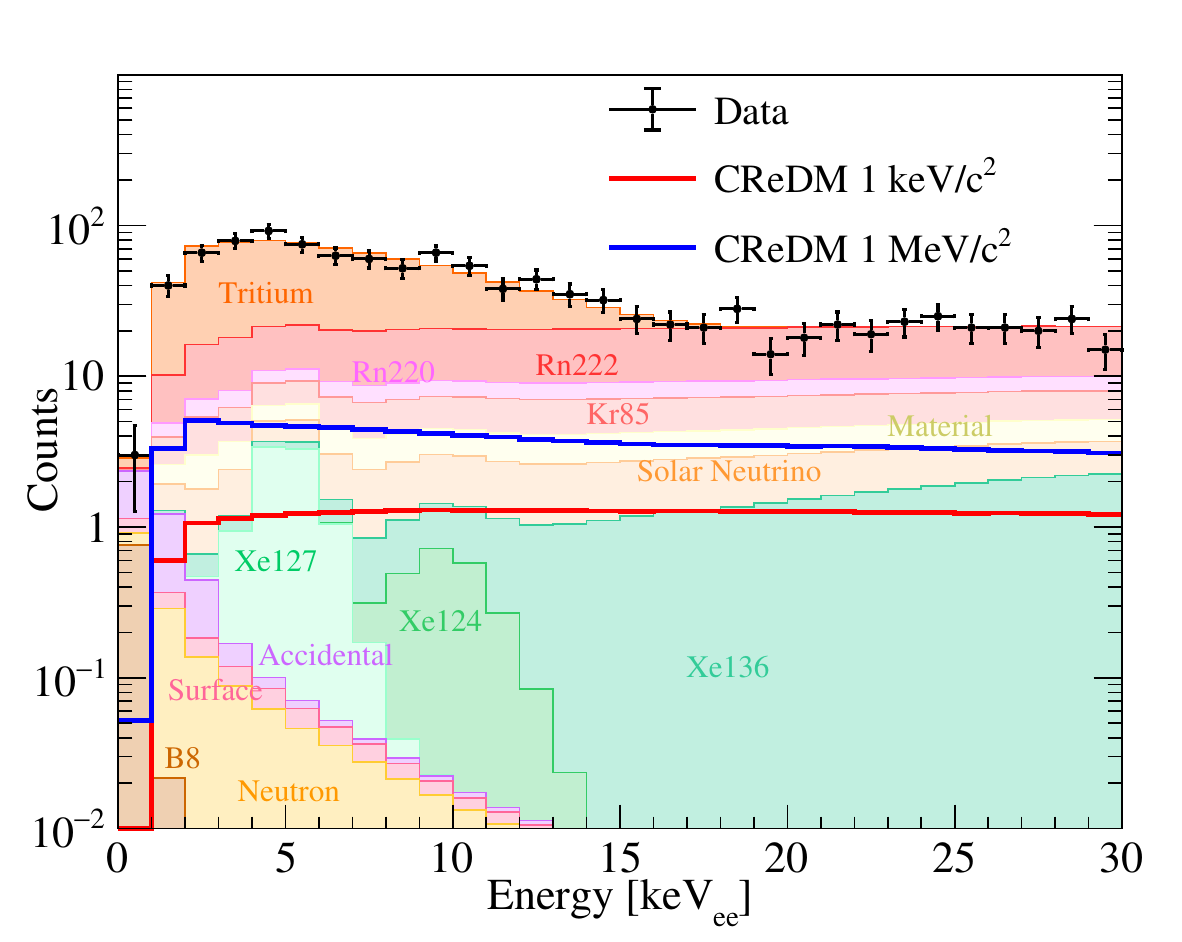}
  \caption{Energy distribution of selected events in data and stacked background components from the background-only fit. The expected signals in PandaX-4T for DM masses of 1 keV/$c^2$ and 1 MeV/$c^2$ are shown in solid red and blue lines (unstacked), with assumed DM-electron cross section of  10$^{-36}$~cm$^2$ and 10$^{-31}$~cm$^2$, respectively.}
  \label{fig:spec_heavy}
\end{figure}

\begin{figure}[h]
  \includegraphics[width=0.48\textwidth]{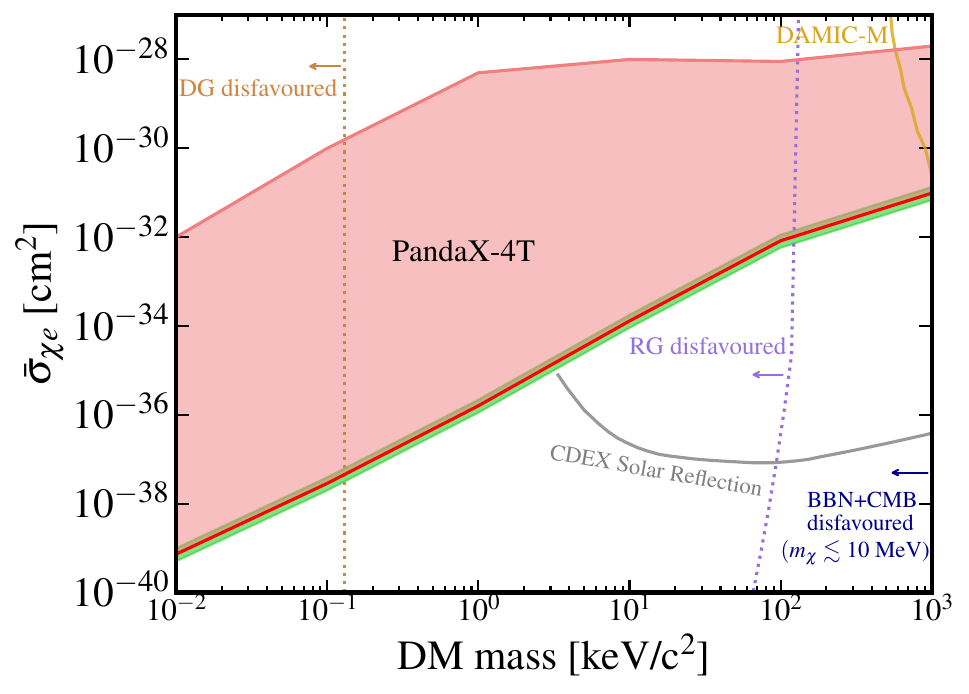}
  \caption{The 90\% CL exclusion region (red region), as well as the green $\pm$1$\sigma$ sensitivity band for the lower boundary. For comparison, the constraints from DAMIC-M~\cite{DAMIC-M:2023gxo} (gold) and CDEX Solar Reflection~\cite{CDEXsolar} (gray) are also shown.
  The brown dotted line indicates the  constraint from Pauli exclusion principle applied to dwarf galaxies (DG)~\cite{Alvey:2020xsk}, below which masses are disfavoured (as indicated by the arrow). Complementary constraints from cosmology and astrophysics are also overlaid, including BBN and CMB~\cite{bbn_constraints,cmb_constraints}, and stellar cooling constraints from red giant (RG) stars~\cite{stellar_cooling}.
  }
  \label{fig:limit}
\end{figure}

A binned likelihood fit is carried out on the data with measured and background spectra.  The background systematic uncertainties have been independently estimated based on Refs.~\cite{PandaX-4T:2021bab,solar_PP}.
Fig.~\ref{fig:spec_heavy} shows the energy distribution in data, in comparison to the background-only best-fit spectra.
After the fit, the data distribution is consistent with no excess of signal.
A profile likelihood ratio (PLR)~\cite{Cowan:2010js,PLR_ref} is constructed as the test statistics to derive the upper limits of CReDM signals. Fig.~\ref{fig:limit} shows the 90\% CL exclusion on the DM-electron cross sections, together with the $\pm1\sigma$ sensitivity band obtained from the background-only pseudo-data. The lowest DM mass in Fig.~\ref{fig:limit} is set to be $10$~eV/$c^2$ to avoid the constraint by the Pauli exclusion principle with the required local DM density for fermionic DM~\cite{jackson2023search}.
For $ m_\chi \ll m_e$, the lower exclusion boundary is approximately proportional to $m_\chi^2$ for the following reasons. The accelerated CReDM flux ${d\Phi}/{dT_\chi}$ roughly scales with 1/$m_\chi^2$ due to the DM number density and that $d\sigma/dT_\chi$ approximately scales with $m_\chi/\mu^2_{\chi e}\sim 1/m_\chi$ (see Eq.~\eqref{eq:diff_flux}). The differential cross section in the detector in Eq.~\eqref{eq:xsec_bound} also scales with $1/\mu^2_{\chi e}\sim 1/m_\chi^2$. The resulting $1/m_\chi^4$ dependence of the expected number of signal events leads to a $m_\chi^2$ dependence on the lower exclusion boundary of $\bar{\sigma}_{\chi e}$.
The earth shielding effects drive the upper exclusion boundary, above which DM particles can barely reach the detector. 
Near the minimum $T_\chi$ required for the detection threshold ($\sim$ 1 keV), the average energy loss of DM per scattering has an increasing dependence on $m_\chi$. This results in stronger Earth attenuation and decreases the slope of the upper boundary for $m_\chi$ $>$ 1~keV/$c^2$.
The theoretical uncertainties from CR distribution and DM density profile are estimated to introduce a variation of 12\% or less on the lower exclusion boundary, by comparing results from an assumed homogeneous CR distribution with local interstellar spectrum and various DM profiles~\cite{ISOthermal,Einasto}.

As shown in Fig.~\ref{fig:limit}, the direct-detection experimental coverage for the fermionic DM-electron interaction in the sub-MeV range is sparse.
In comparison to the recent CDEX constraint using solar reflection~\cite{CDEXsolar}, we extend the mass range down by more than two orders of magnitude.
There are other stringent constraints on sub-MeV DM obtained from astrophysical and cosmological observations shown in the figure, such as stellar cooling~\cite{stellar_cooling} and big bang nucleosynthesis (BBN) combined with cosmic microwave background (CMB)~\cite{bbn_constraints,cmb_constraints}. In addition, the Pauli exclusion principle applied to dwarf galaxies can derive lower bounds on the fermionic DM mass around $\mathcal{O}(0.1)~\text{keV}$~\cite{Alvey:2020xsk}. Our work, on the other hand, derives limits based on experimental data and relies only on information of the present-day local Universe, therefore is highly complementary to the astrophysical and cosmological constraints.

In summary, we presented a new search for the DM-electron interactions from the PandaX-4T detector at CJPL. This is the first time that a direct-detection experiment searches for DM boosted by electrons in cosmic rays, complementary to our previous search for DM boosted by nuclei. Neutrino experiments may also be sensitive to the same parameter region using this boosted DM scenario~\cite{CReDM_2018}. With the same tree-level vector DM-electron interaction in the DM boosting, attenuation, and detection, the search strategy bears little theoretical uncertainty in the fundamental DM properties. In addition, the relativistic effects are properly and consistently evaluated in the acceleration and attenuation processes. With a 0.63 tonne$\cdot$year exposure, no excess of events are observed above the background prediction. We present constraints on DM-electron cross sections for the unexplored DM mass in direct detection experiments between 10~eV/$c^2$ and 3~keV/$c^2$, pushing the current mass range down by more than 2 orders of magnitude, all the way to the minimum fermionic particle mass allowed by the local density of DM.

\input{acknowledgement}

\bibliography{apssamp}

\end{document}

%% file: authorlist.tex

\def\shKeyLab{School of Physics and Astronomy, Shanghai Jiao Tong University, Key Laboratory for Particle Astrophysics and Cosmology (MoE), Shanghai Key Laboratory for Particle Physics and Cosmology, Shanghai 200240, China}
\def\scKeyLab{Jinping Deep Underground Frontier Science and Dark Matter Key Laboratory of Sichuan Province, Liangshan 615000, China}
\def\BUAA{School of Physics, Beihang University, Beijing 102206, China}
\def\BUAACenter{Peng Huanwu Collaborative Center for Research and Education, Beihang University, Beijing 100191, China}
\def\BUAALab{Beijing Key Laboratory of Advanced Nuclear Materials and Physics, Beihang University, Beijing, 102206, China}
\def\SCNT{Southern Center for Nuclear-Science Theory (SCNT), Institute of Modern Physics, Chinese Academy of Sciences, Huizhou 516000, China}
\def\USTClab{State Key Laboratory of Particle Detection and Electronics, University of Science and Technology of China, Hefei 230026, China}
\def\USTCdep{Department of Modern Physics, University of Science and Technology of China, Hefei 230026, China}
\def\BUAALab{International Research Center for Nuclei and Particles in the Cosmos \& Beijing Key Laboratory of Advanced Nuclear Materials and Physics, Beihang University, Beijing 100191, China}
\def\pku{School of Physics, Peking University, Beijing 100871, China}
\def\YaLongSD{Yalong River Hydropower Development Company, Ltd., 288 Shuanglin Road, Chengdu 610051, China}
\def\IAP{Shanghai Institute of Applied Physics, Chinese Academy of Sciences, 201800 Shanghai, China}
\def\CHEPpku{Center for High Energy Physics, Peking University, Beijing 100871, China}
\def\SDUdep{Research Center for Particle Science and Technology, Institute of Frontier and Interdisciplinary Science, Shandong University, Qingdao 266237, Shandong, China}
\def\SDUlab{Key Laboratory of Particle Physics and Particle Irradiation of Ministry of Education, Shandong University, Qingdao 266237, Shandong, China}
\def\UMD{Department of Physics, University of Maryland, College Park, Maryland 20742, USA}
\def\TDLee{New Cornerstone Science Laboratory, Tsung-Dao Lee Institute, Shanghai Jiao Tong University, Shanghai, 200240, China}
\def\MESJTU{School of Mechanical Engineering, Shanghai Jiao Tong University, Shanghai 200240, China}
\def\SYU{School of Physics, Sun Yat-Sen University, Guangzhou 510275, China}
\def\SYUSFI{Sino-French Institute of Nuclear Engineering and Technology, Sun Yat-Sen University, Zhuhai, 519082, China}
\def\NKU{School of Physics, Nankai University, Tianjin 300071, China}
\def\YTU{Department of Physics,Yantai University, Yantai 264005, China}
\def\FDU{Key Laboratory of Nuclear Physics and Ion-beam Application (MOE), Institute of Modern Physics, Fudan University, Shanghai 200433, China}
\def\USST{School of Medical Instrument and Food Engineering, University of Shanghai for Science and Technology, Shanghai 200093, China}
\def\SJTUSC{Shanghai Jiao Tong University Sichuan Research Institute, Chengdu 610213, China}
\def\SPEIT{SJTU Paris Elite Institute of Technology, Shanghai Jiao Tong University, Shanghai, 200240, China}
\def\NNU{School of Physics and Technology, Nanjing Normal University, Nanjing 210023, China}
\def\SYSUzhuhai{School of Physics and Astronomy, Sun Yat-Sen University, Zhuhai, 519082, China}

\affiliation{\TDLee}
\author{Xiaofeng Shang}\affiliation{\shKeyLab}
\author{Abdusalam Abdukerim}\affiliation{\shKeyLab}
\author{Zihao Bo}\affiliation{\shKeyLab}
\author{Wei Chen}\affiliation{\shKeyLab}
\author{Xun Chen}\affiliation{\TDLee}\affiliation{\shKeyLab}\affiliation{\SJTUSC}\affiliation{\scKeyLab}
\author{Yunhua Chen}\affiliation{\YaLongSD}\affiliation{\scKeyLab}
\author{Chen Cheng}\affiliation{\SYU}
\author{Zhaokan Cheng}\affiliation{\SYUSFI}
\author{Xiangyi Cui}\email[Corresponding author: ]{hongloumeng@sjtu.edu.cn}\affiliation{\TDLee}
\author{Yingjie Fan}\affiliation{\YTU}
\author{Deqing Fang}\affiliation{\FDU}
\author{Lisheng Geng}\affiliation{\BUAA}\affiliation{\BUAACenter}\affiliation{\BUAALab}\affiliation{\SCNT}
\author{Karl Giboni}\affiliation{\shKeyLab}\affiliation{\scKeyLab}
\author{Xuyuan Guo}\affiliation{\YaLongSD}\affiliation{\scKeyLab}
\author{Chencheng Han}\affiliation{\TDLee} 
\author{Ke Han}\affiliation{\shKeyLab}\affiliation{\scKeyLab}
\author{Changda He}\affiliation{\shKeyLab}
\author{Jinrong He}\affiliation{\YaLongSD}
\author{Di Huang}\affiliation{\shKeyLab}
\author{Junting Huang}\affiliation{\shKeyLab}\affiliation{\scKeyLab}
\author{Zhou Huang}\affiliation{\shKeyLab}
\author{Ruquan Hou}\affiliation{\SJTUSC}\affiliation{\scKeyLab}
\author{Yu Hou}\affiliation{\MESJTU}
\author{Xiangdong Ji}\affiliation{\UMD}
\author{Xiangpan Ji}\affiliation{\NKU}
\author{Yonglin Ju}\affiliation{\MESJTU}\affiliation{\scKeyLab}
\author{Chenxiang Li}\affiliation{\shKeyLab}
\author{Jiafu Li}\affiliation{\SYU}
\author{Mingchuan Li}\affiliation{\YaLongSD}\affiliation{\scKeyLab}
\author{Shuaijie Li}\affiliation{\YaLongSD}\affiliation{\shKeyLab}\affiliation{\scKeyLab}
\author{Tao Li}\affiliation{\SYUSFI}
\author{Qing Lin}\affiliation{\USTClab}\affiliation{\USTCdep}
\author{Jianglai Liu}\email[Spokesperson: ]{jianglai.liu@sjtu.edu.cn}\affiliation{\TDLee}\affiliation{\shKeyLab}\affiliation{\SJTUSC}\affiliation{\scKeyLab}
\author{Congcong Lu}\affiliation{\MESJTU}
\author{Xiaoying Lu}\affiliation{\SDUdep}\affiliation{\SDUlab}
\author{Lingyin Luo}\affiliation{\pku}
\author{Yunyang Luo}\affiliation{\USTCdep}
\author{Wenbo Ma}\affiliation{\shKeyLab}
\author{Yugang Ma}\affiliation{\FDU}
\author{Yajun Mao}\affiliation{\pku}
\author{Yue Meng}\affiliation{\shKeyLab}\affiliation{\SJTUSC}\affiliation{\scKeyLab}
\author{Xuyang Ning}\affiliation{\shKeyLab}
\author{Binyu Pang}\affiliation{\SDUdep}\affiliation{\SDUlab}
\author{Ningchun Qi}\affiliation{\YaLongSD}\affiliation{\scKeyLab}
\author{Zhicheng Qian}\affiliation{\shKeyLab}
\author{Xiangxiang Ren}\affiliation{\SDUdep}\affiliation{\SDUlab}
\author{Nasir Shaheed}\affiliation{\SDUdep}\affiliation{\SDUlab}
\author{Xiyuan Shao}\affiliation{\NKU}
\author{Guofang Shen}\affiliation{\BUAA}
\author{Manbin Shen}\affiliation{\YaLongSD}\affiliation{\scKeyLab}
\author{Lin Si}\affiliation{\shKeyLab}
\author{Wenliang Sun}\affiliation{\YaLongSD}\affiliation{\scKeyLab}
\author{Yi Tao}\affiliation{\shKeyLab}\affiliation{\SJTUSC}
\author{Anqing Wang}\affiliation{\SDUdep}\affiliation{\SDUlab}
\author{Meng Wang}\affiliation{\SDUdep}\affiliation{\SDUlab}
\author{Qiuhong Wang}\affiliation{\FDU}
\author{Shaobo Wang}\affiliation{\shKeyLab}\affiliation{\SPEIT}\affiliation{\scKeyLab}
\author{Siguang Wang}\affiliation{\pku}
\author{Wei Wang}\affiliation{\SYUSFI}\affiliation{\SYU}
\author{Xiuli Wang}\affiliation{\MESJTU}
\author{Xu Wang}\affiliation{\TDLee}
\author{Zhou Wang}\affiliation{\TDLee}\affiliation{\shKeyLab}\affiliation{\SJTUSC}\affiliation{\scKeyLab}
\author{Yuehuan Wei}\affiliation{\SYUSFI}
\author{Mengmeng Wu}\affiliation{\SYU}
\author{Weihao Wu}\affiliation{\shKeyLab}\affiliation{\scKeyLab}
\author{Yuan Wu}\affiliation{\shKeyLab}
\author{Mengjiao Xiao}\affiliation{\shKeyLab}
\author{Xiang Xiao}\affiliation{\SYU}
\author{Kaizhi Xiong}\affiliation{\YaLongSD}\affiliation{\scKeyLab}
\author{Binbin Yan}\affiliation{\TDLee}
\author{Xiyu Yan}\affiliation{\SYSUzhuhai}
\author{Yong Yang}\email[Corresponding author: ]{yong.yang@sjtu.edu.cn}\affiliation{\shKeyLab}
\author{Chunxu Yu}\affiliation{\NKU}
\author{Ying Yuan}\affiliation{\shKeyLab}
\author{Zhe Yuan}\affiliation{\FDU} 
\author{Youhui Yun}\affiliation{\shKeyLab}
\author{Xinning Zeng}\affiliation{\shKeyLab}
\author{Minzhen Zhang}\affiliation{\TDLee}
\author{Peng Zhang}\affiliation{\YaLongSD}\affiliation{\scKeyLab}
\author{Shibo Zhang}\affiliation{\TDLee}
\author{Shu Zhang}\affiliation{\SYU}
\author{Tao Zhang}\affiliation{\TDLee}\affiliation{\shKeyLab}\affiliation{\SJTUSC}\affiliation{\scKeyLab}
\author{Wei Zhang}\affiliation{\TDLee}
\author{Yang Zhang}\affiliation{\SDUdep}\affiliation{\SDUlab}
\author{Yingxin Zhang}\affiliation{\SDUdep}\affiliation{\SDUlab} 
\author{Yuanyuan Zhang}\affiliation{\TDLee}
\author{Li Zhao}\affiliation{\TDLee}\affiliation{\shKeyLab}\affiliation{\scKeyLab}
\author{Jifang Zhou}\affiliation{\YaLongSD}\affiliation{\scKeyLab}
\author{Ning Zhou}\affiliation{\TDLee}\affiliation{\shKeyLab}\affiliation{\SJTUSC}\affiliation{\scKeyLab}
\author{Xiaopeng Zhou}\affiliation{\BUAA}
\author{Yubo Zhou}\affiliation{\shKeyLab}
\author{Zhizhen Zhou}\affiliation{\shKeyLab}
\collaboration{PandaX Collaboration}
\author{Shao-Feng Ge}\affiliation{\TDLee}\affiliation{\shKeyLab}
\author{Chen Xia}\email[Corresponding author: ]{xiachen@sjtu.edu.cn}\affiliation{\TDLee}\affiliation{\shKeyLab}
\noaffiliation

%% file: acknowledgement.tex

 
We thank Jie Sheng, Chuan-Yang Xing, and Ran Ding for their helpful discussions. 
This project is supported in part by grants from the National Natural Science Foundation of China (Nos. 12090060, 12090061, 
12090064, 12375101, 
12247148, 12247141), 
a grant from the Ministry of Science and Technology of China (No. 2023YFA1606200), and by Office of Science and Technology, Shanghai Municipal Government (grant No. 22JC1410100). We thank for the support from Double First Class Plan of the Shanghai Jiao Tong University. We also thank the sponsorship from the Chinese Academy of Sciences Center for Excellence in Particle Physics (CCEPP), Hongwen Foundation in Hong Kong, Tencent Foundation in China, and Yangyang Development Fund. Finally, we thank the CJPL administration and the Yalong River Hydropower Development Company Ltd. for indispensable logistical support and other help.